\documentclass[aps,prl,twocolumn,groupedaddress,amsmat,amssymb,amsfonts,superscriptaddress]{revtex4-2}

\usepackage{graphicx}
\usepackage{amsmath,amsfonts,amsthm} 
\usepackage{blkarray}
\usepackage{physics}
\usepackage{mhchem}
\usepackage{caption}
\usepackage[dvipsnames,svgnames]{xcolor}
\usepackage{float}
\usepackage{siunitx}

\usepackage{hyperref}
\hypersetup{
    colorlinks=true,
    citecolor=Green,
    linkcolor=Red,
    urlcolor=NavyBlue
}

\begin{document}
\raggedbottom
\setlength{\abovedisplayskip}{5pt}
\setlength{\belowdisplayskip}{5pt}
\setlength{\abovedisplayshortskip}{0pt}
\setlength{\belowdisplayshortskip}{0pt}

\title{Microscopic Roadmap to a Yao-Lee Spin-Orbital Liquid}
\author{Derek Churchill}
\affiliation{Department of Physics, University of Toronto, Ontario, Canada M5S 1A7}
\author{Emily Z. Zhang}
\affiliation{Department of Physics, University of Toronto, Ontario, Canada M5S 1A7}
\author{Hae-Young Kee}
\email[]{hy.kee@utoronto.ca}
\affiliation{Department of Physics, University of Toronto, Ontario, Canada M5S 1A7}
\affiliation{Canadian Institute for Advanced Research, CIFAR Program in Quantum Materials, Toronto, Ontario, Canada, M5G 1M1}
\date{\today}

\begin{abstract}
The exactly solvable spin-1/2 Kitaev model on a honeycomb lattice has drawn significant interest, as it offers a pathway to realizing the long-sought after quantum spin liquid. Building upon the Kitaev model, Yao and Lee introduced another exactly solvable model on an unusual star lattice featuring non-abelian spinons. The additional pseudospin degrees of freedom in this model could provide greater stability against perturbations, making this model appealing. However, 
a mechanism to realize such an interaction in a standard honeycomb lattice remains unknown.
Here we provide a microscopic theory to obtain the Yao-Lee model on a honeycomb lattice by utilizing strong spin-orbit coupling of anions edge-shared between two $e_g$ ions in the exchange processes. This mechanism leads to the desired bond-dependent interaction among spins rather than orbitals, unique to our model, implying that the orbitals fractionalize into gapless Majorana fermions and fermionic octupolar excitations emerge. Since the conventional Kugel-Khomskii interaction also appears, we examine the phase diagram including these interactions using classical Monte Carlo simulations and exact diagonalization techniques. Our findings reveal a broad region of disordered states that break rotational symmetry in the bond energy, suggesting intriguing behavior reminiscent of a spin-orbital liquid. 
\end{abstract}
\maketitle

\begin{figure}[h]
    \includegraphics[width=0.4\textwidth]{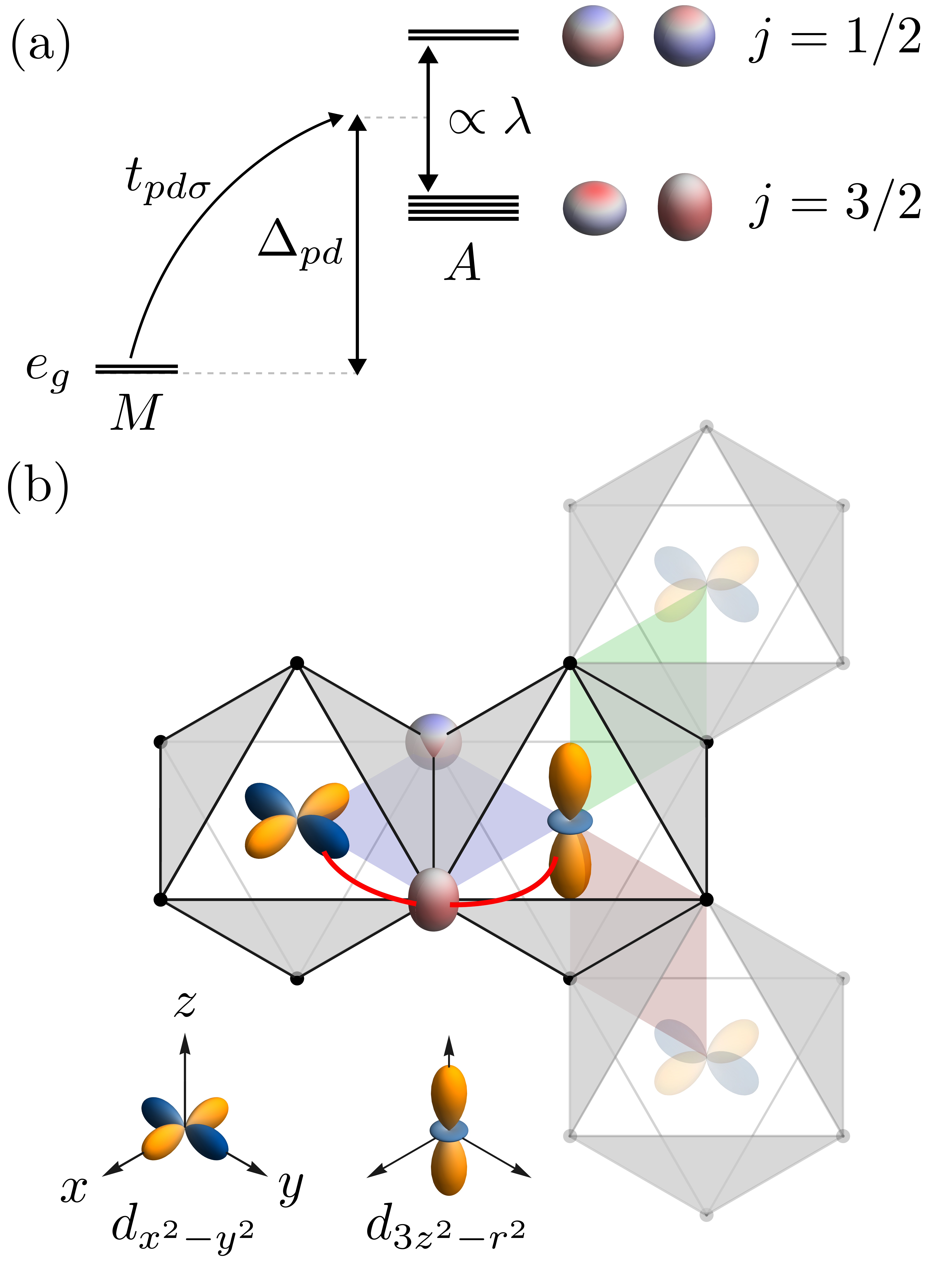}
    \caption{\label{fig1} (a) Virtual hopping process of one hole (or electron) between adjacent $M$ and $A$ sites separated by the charge transfer gap $\Delta_{pd}$. Large SOC $\lambda$ on the $A$ site splits the $p$ orbitals into $j=1/2$ and $j=3/2$ states and is responsible for generating an imaginary spin-dependent hopping. Separation of the $d$ orbital manifold into $t_{2g}$ and $e_g$ irreducible representations due to cubic crystal field splitting and filled $t_{2g}$ are imposed.  Charge densities for the $j=1/2$ Kramer's doublet is shown alongside its label, with the red and blue texture representing spin up and spin down density respectively. Two of the $j=3/2$ states are plotted alongside their label for $j^z=1/2$ (left) and $j^z=3/2$ (right). (b) Honeycomb lattice formed from $M$ sites, located at the center of each octahedral cage of $A$ sites. Orbitals are centered at the atomic sites; positive and negative lobes are encoded by orange and blue respectively. The $x$ (red), $y$ (green) and $z$ (blue) bonds are represented as transparent planes and are related by the honeycomb $C_3$ symmetry. Red lines represent a hole or electron hopping ($\propto t_{pd\sigma}$) between $M$ and $A$ sites; together, these paths provide an inter-orbital bridge via the ligand between sites $M_i$ and $M_j$. Only these two paths are shown for simplicity.}
\end{figure}

\section{Introduction}
Frustration has been routinely shown to lead to remarkable physics. Quantum spin liquids are one consequence, characterized by fractionalized excitations, long-range entanglement and an absence of magnetic ordering down to zero temperature.
\cite{anderson1973MRB,balents2010Nat,Savary2017IOP,Zhou2017RMP,Wen2017RMP,Hermanns2018ARP,Knolle2019ARP,Takagi2019NRP,Broholm2020Sci,Kivelson2023Natrev} 
For many years there has been an extensive search for materials hosting a quantum spin liquid. 
Transition-metal Mott insulators have been particularly fruitful in realizing frustrated interactions due to strong Coulomb interactions, spin-orbit coupling (SOC), and the multi-orbital nature of these compounds.\cite{Khal2005PTPS, Jackeli2009PRL, Rau2014PRL,Krempa2014ARCMP,rau2016ARP,Winter2016PRB,Motome2020,Takayama2021JPSJ,Trebst2022,Rouso2024RoPP}

Kitaev's exactly solvable model\cite{Kitaev2006AOP} is an exemplar of bond-frustration, featuring competing spin-1/2 Ising interactions on the honeycomb lattice, expressed as
\begin{align}
   H_{\textrm{K}} = K \sum_{<ij>_\gamma} \left(S^\gamma_i \cdot S^\gamma_j\right).
\end{align}
where $\gamma=x,y,z$ is the nearest neighbor bond $\langle ij \rangle_\gamma$ label, illustrated in Fig. \ref{fig1}.
The model exhibits a quantum spin liquid ground state with anyonic excitations and has been proposed as a way to achieve fault tolerant quantum computing.\cite{Kitaev2003AOP,Kitaev2006AOP} The main ingredient to generate the Kitaev interaction is through strong SOC at the transition metal site,
which binds spin and orbital degrees of freedom and permits 
bond-dependent spin-orbit entangled pseudospin interactions.\cite{Jackeli2009PRL,Chaloupka2010PRL,Rau2014PRL,rau2016ARP,Liu2018PRB,Liu2020PRL,Motome2020JOP,trebst2022PR,Liu2023PRB}

Motivated by this seminal work, Yao and Lee proposed a generalized version of the Kitaev model, expressed as\cite{Yao2011PRL}
\begin{align}
    H_{\textrm{yl}} = J_{\rm yl} \sum_{<ij>_\gamma} \bigl({\bf S}_i \cdot {\bf S}_j\bigr) \otimes \bigl(T_i^\gamma \cdot T^\gamma_j\bigr),
\end{align}
where 
${\bf S}$ and ${\bf T}$ represent spin and pseudospin, respectively. 
The model emerges as a low-energy effective Hamiltonian on the decorated honeycomb lattice, whose ground state is a Z$_2$ spin liquid.
Yao and Lee solved the model by decomposing the spin and pseudospin into two Majorana fermions, resulting in three species of gapless Majorana fermions coupled to a $\mathbb{Z}_2$ gauge field.
By Lieb's theorem, the ground state lies in the zero flux sector, yielding a Dirac dispersion for the Majorana fermions. 
It exhibits spin-1 fermionic excitations, a zoo of vison crystal phases, and in the presence of an external field, non-abelian spinons.\cite{Yao2011PRL,Chulliparambil2021PRB,Akram2023PRB} Due to the additional degree of freedom, this model is robust to more perturbations than the original Kitaev model \cite{Seifert2020PRL,Chulliparambil2020PRB,Nica2023NPJ,Poliakov2024Arxiv,wu2024Arxiv}, making it an attractive alternative candidate.

While fascinating, identifying the microscopic mechanisms that give rise to such interactions remains a significant challenge. When mapped onto a honeycomb lattice, the additional degrees of freedom may arise from orbital fluctuations. However, it is well-established that when orbital fluctuations play a role in exchange processes, models like the Kugel-Khomskii (KK) model, characterized by a product of spin-Heisenberg and orbital-Heisenberg interactions\cite{Kliment1982Sov}, or the compass model\cite{Nussinov2015RMP} typically emerge. This raises the question of how to generate bond-dependent Kitaev interactions acting on either spins or orbitals, while retaining the Heisenberg interaction for the other.

To appreciate the difficulty inherent in this problem, it is essential to recognize that the Kitaev interaction requires orbital changes through hopping, which depend on the bond direction.\cite{Khal2005PTPS,Jackeli2009PRL,Rau2014PRL,Rouso2024RoPP} Therefore, the origin of the Kitaev interaction lies in the orbitals, which manifest as pseudospin bond-dependent interactions via SOC. If we treat orbitals as a separate degree of freedom from spins, the result is either the KK or compass model. Conversely, introducing the SOC required for Kitaev interactions entangles spin and orbital degrees of freedom, eliminating the possibility of treating them separately, as is necessary for generating the YL model.


Here, we outline how to overcome this challenge and present a roadmap for generating the Yao-Lee (YL) model on a honeycomb lattice. By leveraging the strong SOC of anions in the exchange process within an edge-sharing structure like the honeycomb lattice,  we find a YL-type spin-orbital model.  The key idea is that the spin-orbital model obtained through indirect exchange — via an intermediate {\it anion} with strong SOC — between neighboring e$_g$ ions includes the YL interaction, along with a weaker KK interaction.
For the 90$^\circ$ edge sharing lattice such as the honeycomb, interorbital hopping between $e_g$ orbitals is only allowed via the anion, making this exchange process dominate over intraorbital direct exchange. 

Utilizing classical Monte Carlo (MC) simulations and exact diagonalization (ED), we map the classical and quantum phase diagrams, revealing a large disordered region that breaks rotational symmetry and encompasses an exactly solvable YL spin-orbital liquid. Our work offers a microscopic pathway to transition from the traditional KK limit to the bond-dependent YL limit in orbital-fluctuating Mott insulators. We propose that $d^9$ or low-spin $d^7$ systems, surrounded by heavy ions forming a honeycomb lattice, would serve as promising candidate materials.

\section{Result}
\subsection{Microscopic derivation }
We first recall how the KK model arises to make a comparison with the mechanism of the YL interaction.
Consider one electron or hole in two degenerate orbitals at $M$ sites, leading to fluctuations between the two orbitals. 
To represent the orbitals, we introduce pseudospin-1/2 operators $T^k=\frac{1}{2} \sigma^k$ with $k=x,y,z$, where $\sigma^k$ are Pauli matrices, such that $|a\rangle$ and $|b\rangle$ are eigenstates of $T^z$ with eigenvalues $\pm \frac{1}{2}$ respectively. Orbital ($m$) preserving, intraorbital hopping of electrons between neighbouring transition metal ($M$) sites $M_i$ and $M_j$,
\begin{align}
    t^{\textrm{direct}}_{ij} = \sum_{m, \sigma} t_m c^{\dagger}_{i, m \sigma} c_{j, m \sigma} + \textrm{H.c.},
\end{align}
where $t_m$ is the hopping amplitude, results in the low-energy spin and orbital model colloquially known as the KK model \cite{Kliment1982Sov,vandb2001PRB,Mostovoy2002PRL,Khomskii2016JETP}.
The SU(4) Kugel-Khomskii model is a classic example of frustration between degrees of freedom and is given by
\begin{align}
    H_{\textrm{KK}} = J_{\rm kk} \sum_{\langle ij \rangle} \left({\bf S}_i \cdot {\bf S}_j + \frac{1}{4}\right) \otimes \left({\bf T}_i \cdot {\bf T}_j + \frac{1}{4}\right).
\end{align}
where $J_{\rm kk} = \frac{8 t_m^2}{U}$ when the direct hopping $t_m$ is the same for the two orbitals. It is characterized by isotropic interactions between nearest-neighbour spin-1/2 ($S_i$) and pseudospin-1/2 ($T_i$) degrees of freedom, and has been suggested to display a spin-liquid ground state on the honeycomb lattice\cite{Corboz2012PRX,Jakab2016PRB,Yamada2018PRL,Natori2019PRB,Jin2023PRB,Vordos2023Arxiv}, but it lacks exact solvability. 

At this stage, exchange between nearest neighbour $M$ sites will generate pseudospin compass and KK exchange interactions depending on the lattice geometry and relative hopping amplitudes. \cite{Kliment1982Sov,Nussinov2015RMP} The Kitaev interaction is absent. The key missing element is SOC, which enables bond-dependent spin interactions, while allowing the other degrees of freedom to behave in a Heisenberg fashion.

Now let us consider 
a honeycomb lattice of $e_g$ orbitals at $M$ sites, surrounded by edge-shared octahedra of heavy anions ($A$) with large SOC ($\lambda$), as illustrated in Fig. 1.
Due to the bonding geometry, the direct interorbital hopping integral is absent and indirect {\it interorbital} hopping via p-orbital superexchange, which mediates a change in angular momentum, dominates over the direct intraorbital hopping.
Large anion SOC on the $A$ site leads to a splitting of the single hole states into $j=1/2$ and $j=3/2$ manifolds, separated by an energy $\frac{3}{2}\lambda$ as shown in Fig. \ref{fig1}. 

After taking into account the SOC on the $p$-orbitals at $A$ sites, we determine an effective nearest-neighbour hopping between $e_g$ ions by integrating out the ligand $p$-orbitals. 
Due to the symmetry imposed by the ideal octahedra and $90^{\circ}$ bonding geometry, seen in Fig. \ref{fig1}, only hopping via $\sigma$-bonds ($t_{pd\sigma}$) between $d$- and $p$-orbitals at $M$ and $A$ sites, respectively, is permitted. Assuming that $t_{pd\sigma}$ is much smaller than the charge-transfer gap $\Delta_{pd}$, the effective hopping between $e_g$ ions at the sites $M_i$ and $M_j$ on a z-bond though the anion p-orbitals is given by,
\begin{align}
    t^z_{ij} = i t_{\textrm{eff}} \sum_{\sigma} (-1)^{\sigma} c^{\dagger}_{i, a \sigma} c_{j, b \sigma} + \textrm{H.c.},
\end{align}
where $t_{\textrm{eff}}= \frac{t_{pd\sigma}^2}{4 \sqrt{3}} \left(\frac{1}{\Delta_{pd}-\frac{\lambda}{2}}-\frac{1}{\Delta_{pd}+\lambda} \right)$, and $\sigma = \pm 1$ for spin up and down respectively. 
We emphasize that this hopping is only finite with SOC present on the $A$ site, which can be seen explicitly in the expression for $t_{\textrm{eff}}$, where an interference between different exchange paths results in an exact cancellation when $\lambda = 0$. 

We determine the exchange interactions using strong-coupling perturbation theory
 truncated at second-order, assuming that $t_{\textrm{eff}}$ is small compared to the onsite Coulomb interaction, $U$. For further details, we refer to the supplementary information (SI)~\cite{SM}.
If the $M$ sites are well-separated such that the dominant exchange process is through the ligands, a strong-coupling expansion 
yields the final form of the effective spin-orbital model
\begin{eqnarray}\label{zbondmodel}
    H_{\rm eff} =& & -J \sum_{\langle ij \rangle_\gamma} \left[ \biggl({\bf S}_i \cdot {\bf S}_j - 2 S_i^\gamma S_j^\gamma - \frac{1}{4}\biggr) \right. \nonumber\\
   && \hspace{1.5cm} \left. \otimes \biggl({\bf T}_i \cdot {\bf T}_j - 2 T_i^y T_j^y - \frac{1}{4}\biggr) \right],
\end{eqnarray}
where $J \equiv \frac{8 t_{\textrm{eff}}^2}{U}$.
The YL interaction with spin-1/2 having Kitaev type,  $(S_i^\gamma S_j^\gamma) ({\bf T}_i \cdot {\bf T}_j)$, is uncovered along with other interactions. Note that the bond-dependent interaction acts on spins, not on orbitals. The implications of this difference from the original Yao-Lee (YL) model will be discussed later.
It is also noteworthy that in the absence of orbital fluctuations, as in the case of 
$d^8$ (i.e., two electrons or holes in the $e_g$ orbital), the spin interaction becomes a sum of Heisenberg and Kitaev interactions, with the Kitaev term being twice as large as the Heisenberg term. This effect resembles the higher-spin Kitaev interaction mechanism described in \cite{Stavro2019prl}, where the factor of 2 difference originates from the $e_g$ wavefunctions.

Taking into account the direct intraorbital hopping process that leads to the KK interaction modifies the coefficients in front of $({\bf S}_i \cdot {\bf S}_j) ({\bf T}_i \cdot {\bf T}_j)$, ${\bf S}_i \cdot {\bf S}_j$, and ${\bf T}_i \cdot{\bf T}_j$.
Note that the direct intraoribital hopping generated KK interaction Eq. (4) has the opposite sign, making the overall KK interaction strength weaker. We introduce two parameters $\alpha$ and $\beta$ to account for the exchange couplings $J$ and $J_{kk}$.
We study the following model Hamiltonian for the $\gamma=x,y,z$ bond to understand the impact of other terms to the YL model:
\begin{eqnarray}\label{fullmodel}
    \notag H_{\rm model}= & &-\sum_{\langle ij \rangle_\gamma} \biggl[\biggl(\alpha {\bf S}_i \cdot {\bf S}_j - 2 S_i^\gamma S_j^\gamma - \beta\biggr) \\
    && \hspace{1.2cm} \otimes \biggl(\tilde{{\bf T}}_i \cdot \tilde{{\bf T}}_j - \beta\biggr)\biggr].
\end{eqnarray} 
$H_{\rm eff}$ corresponds to $\alpha =1$ and $\beta = 1/4$, while the YL and KK limit correspond to $\alpha = \beta =0$ and $\alpha \rightarrow \infty$, respectively. 
Here we also introduced $\tilde{T_i}$ denoting the following sublattice transformation on the pseudospin operators,
\begin{align}\label{sublat}
    T_{i}^x \rightarrow \tilde{T}_{i}^x, \quad T_{i}^y \rightarrow (-1)^i\tilde{T}_{i}^y, \quad T_{i}^z \rightarrow \tilde{T}_{i}^z,
\end{align}
to cast it into Heisenberg form for simplicity. 
This does not change the phase diagram, but its impact on the Majorana excitations is discussed later.

Below we 
present the classical and quantum phase diagrams,  with $J\equiv 1$, by tuning $\alpha$ and $\beta$  to understand how the YL limit is connected to the KK limit followed by the MC and ED methods used to obtain the phase diagrams.

\subsection{ Classical Phase Diagram}
\begin{figure}[h]
    \includegraphics[width=0.42\textwidth]{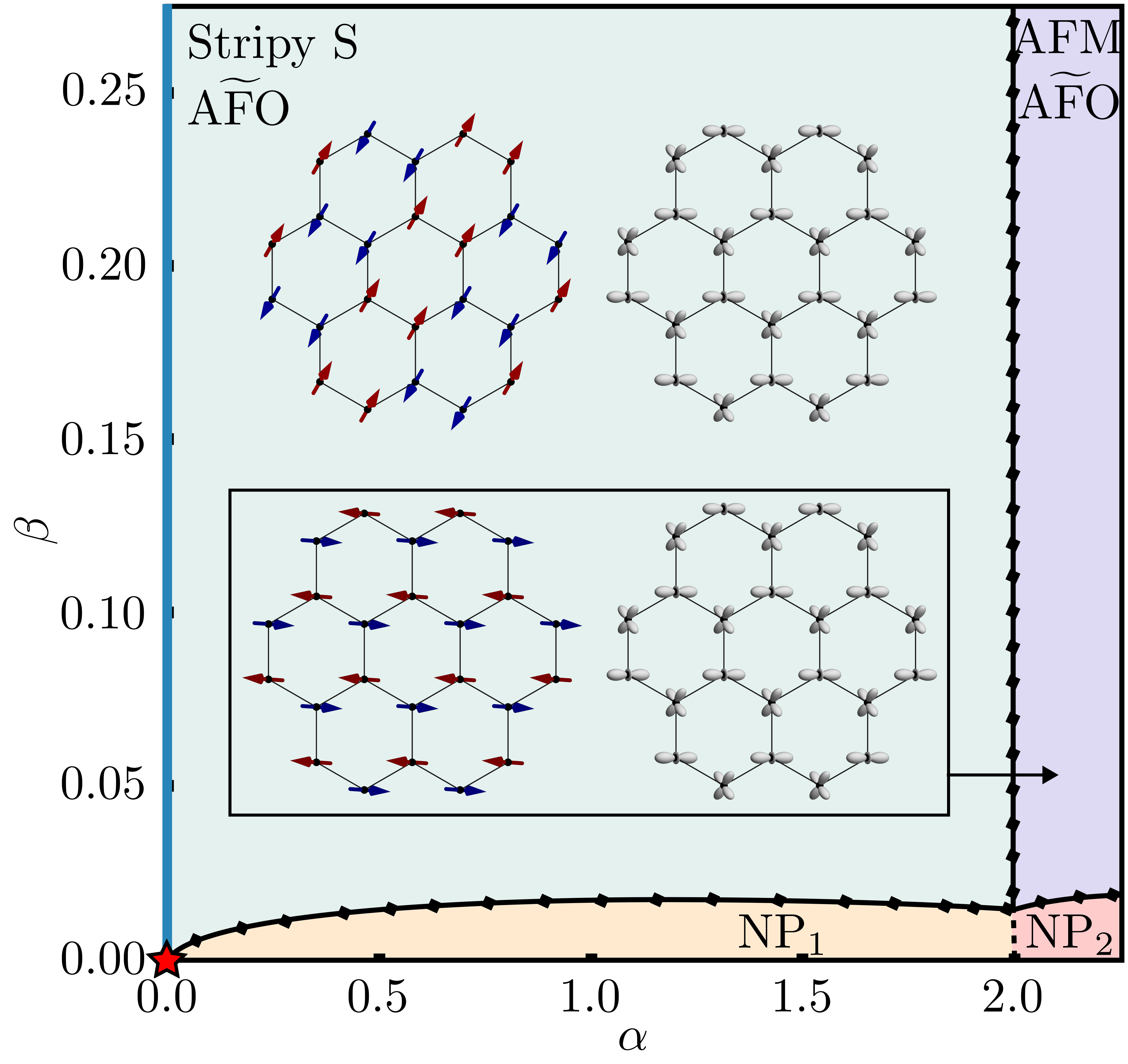}
    \caption{\label{fig2} Classical phase diagram determined using Monte Carlo simulations. Solid black lines denote interpolated phase boundaries and square markers denote a small subset of points where the second derivative of the ground state energy was calculated and diverges. Stripy spin / $\tilde{AFO}$ and AFM / $\tilde{AFO}$ orders are shown in the insets. Arrows which are directed in the $+\hat{c}$ ($-\hat{c}$) direction are coloured red (blue), and black arrows lie within the $ab$ plane. Classical spin/orbital configurations and correlators for nematic phases (NP$_1$ and NP$_2$) are shown in the SI. We use a thick blue line to emphasize the spin disordered line along $\alpha=0$, and a red star to represent the classical YL model.}
\end{figure}

The classical phase diagram of Eq. \ref{fullmodel} on a honeycomb lattice as a function of $\alpha$ and $\beta$ is presented in Fig. \ref{fig2}. We perform classical Monte Carlo simuations by treating the spin and orbital degrees of freedom as vectors with magnitude $\frac{1}{2}$ in $\mathbb{R}^3$. We take into account both spins and orbitals by considering a honeycomb bilayer with each layer representing its own degree of freedom, coupled by the quartic interactions in Eq. \ref{fullmodel}. Details of the cluster geometry and simulation parameters can be found in the SI~\cite{SM}.

Let us first discuss the large $\beta$ regime.
For large $\beta$, we find two ordered phases depending on the value of $\alpha$. The structure of the ordered phases bears a strong resemblance to the phase diagram of the pure $JK$ model on the honeycomb lattice.\cite{Chaloupka2010PRL,Rau2014PRL} In particular, when $0 < \alpha < 2$, we find an ordered stripy spin and antiferro-orbital$(\tilde{AFO})$ phase, where the tilde represents the ordering pattern after the $T^y$ sublattice transformation, Eq. (\ref{sublat}). The red and blue arrows in the inset of Fig. \ref{fig2} represent the stripy spin order, while the alternating orbital shapes denote antiferro-orbital order. While there are only three degenerate configurations for the stripy spin configuration, the degeneracy of the classical ground state manifold is vastly expanded by the $SO(3)$ symmetric pseudospin interaction. When $\alpha=2$, there is a second-order transition to the antiferromagnetic (AFM) spin , while the orbital ordering $\tilde{AFO}$ phase remains the same (shown in the bottom inset of Fig. \ref{fig2}), similar to the pure $JK$ model which undergoes a transition from stripy spin to AFM order when $J>0$ and $J=-K$. This transition is insensitive to the value of $\beta$ as indicated by a straight transition line in Fig. \ref{fig2}.

For small $\beta$, we identify  two regions of intriguing phases.
To understand the nature of these phases, we compute the structure factors 
\begin{eqnarray}
    S(Q)&=&\frac{1}{N^2}\sum_{ij} \expval{\left(S_i \cdot S_j\right)} e^{-i Q \cdot (r_i-r_j)}, \nonumber\\
    T(Q)&=&\frac{1}{N^2}\sum_{ij} \expval{\left(T_i \cdot T_j\right)} e^{-i Q \cdot (r_i-r_j)},
\end{eqnarray}
and 
the spin-orbital (SO) correlator given by
\begin{align}
    &\expval{ST(Q)} = \frac{1}{N^2}\sum_{ij} \expval{\left(S_i \cdot S_j\right)\left(T_i \cdot T_j\right)} e^{-i Q \cdot (r_i-r_j)}.
\end{align}
Both phases display no ordering in the spin and orbital correlators. $\langle \sum_{i} {\bf S} _i \rangle =0$, but a higher rank
for instance $\langle \sum_{i} (S^a_i)^2 \rangle \neq 0$. This suggests that a spin acts like a director with a specific orientation chosen ($a$-axis in this case, which is perpendicular to one of the honeycomb bonds) like in the nematic phase. For this reason we will refer them as nematic paramagnetic (NP) phases, denoted by NP$_1$ and NP$_2$. The orbitals follow the spin nematic pattern, and the SO correlators show a peak at $M$ and $\Gamma$ points in NP$_1$ and NP$_2$, respectively. The details of the correlators are shown in the SI.

The line $\alpha = 0$ denoted by the solid blue line in Fig. \ref{fig2} is a region of particular interest. Along this line, when $\beta = 0$, both spins and orbitals are completely disordered, and we represent this point by a red star in Fig. \ref{fig2}, which corresponds to the YL limit. 
As soon as $\beta \neq 0$, the orbitals immediately exhibit $\tilde{AFO}$ order due to the bilinear terms overcoming the frustrated quartic interactions exhibiting a peak in $T(Q)$ at $\Gamma$ point (see SI), while the spin is disordered.
It is interesting that $\alpha$ that effectively includes the Heisenberg interaction does not lead to a conventional spin or orbital ordered phase unlike  $\beta$ which is more detrimental to the disordered YL limit. The plots for the classical orders and structure factors can be found in the SI.

\subsection{ Quantum Phase Diagram} 
\begin{figure}[h]
    \includegraphics[width=0.42\textwidth]{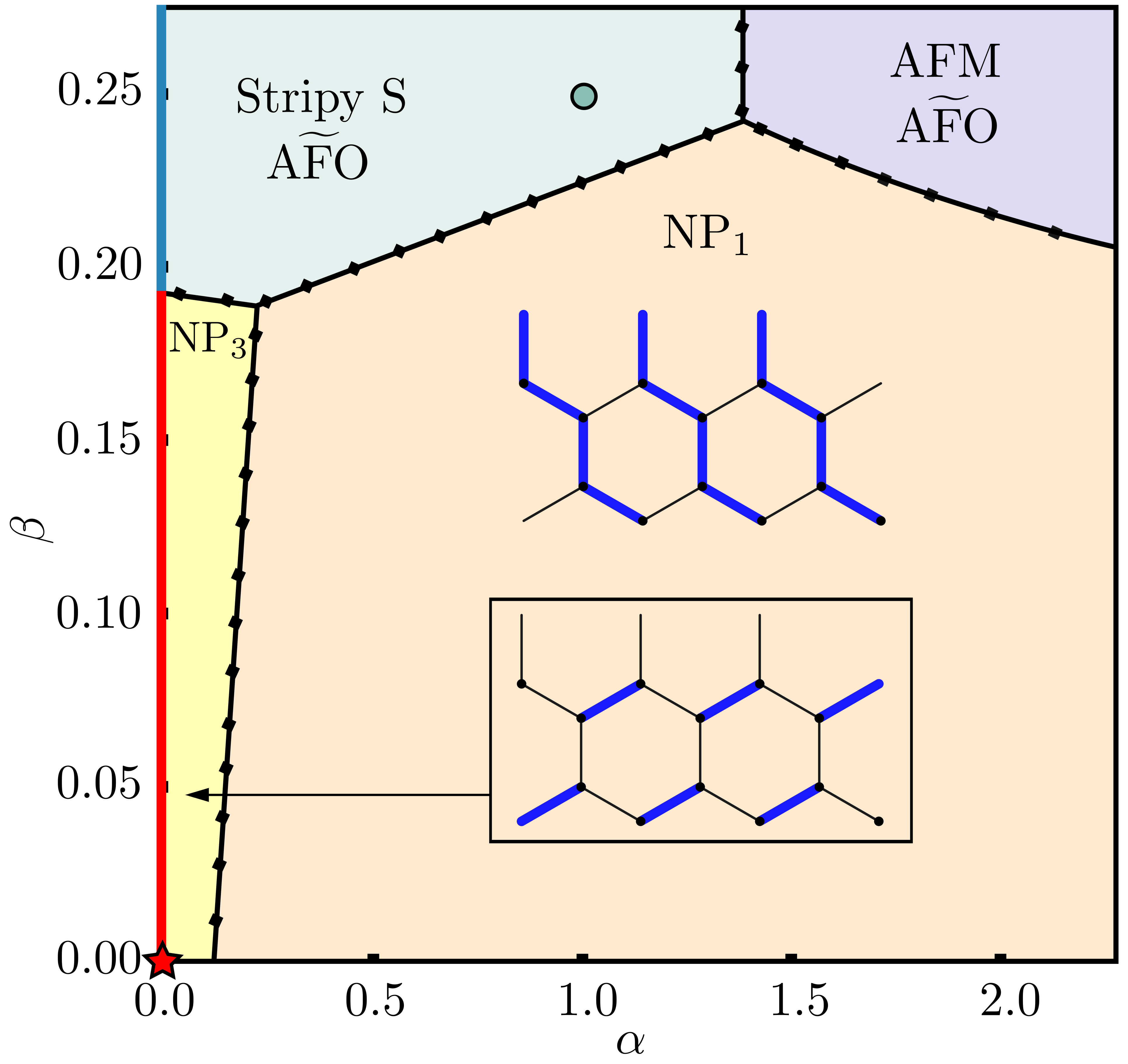}
    \caption{\label{fig3} Quantum phase diagram determined using exact diagonalization. The green dot denotes the ground state of the effective model $H_{\rm eff}$ corresponding to $\alpha=1$ and $\beta=\frac{1}{4}$.  NP$_1$ and NP$_3$ denote the NP phases in the quantum model that were distinguished by looking at bond-energy correlators shown in the inset.  Bonds with lower energy are denoted by dark blue lines. The light blue line at $\alpha=0$ for large $\beta$ denotes a region with no signatures of long-range spin or orbital order. The exactly solvable point is denoted by a red star and exhibits a spin liquid ground state, which extends to $\beta \approx 0.19$ and is emphasized by a thick red line.}
\end{figure}

The quantum phase diagram is presented in Fig. \ref{fig3}, which was determined with the QuSpin exact diagonalization package; simulation details are included in the SI~\cite{SM}. \cite{Weinberg2019SPP}
The quantum phase diagram exhibits the same ordered phases as the classical case with the phase boundaries vastly shifted. 
For large $\beta$ around 1/4, the transition line between stripy and AFM spin with $\tilde{AFO}$ order occurs at a smaller $\alpha$, similar to the $JK$ model. \cite{Chaloupka2010PRL,Rau2014PRL}

The disordered regions in the quantum model have extended to a much larger $\beta$. A wide regime of the phase space is occupied by NP$_1$. As in the classical phase diagram, the disordered NP phases both have vanishing bilinear correlators $S(Q)$ and $T(Q)$, indicating disorder. However, these orders spontaneously break the lattice $C_3$ symmetry, which we show by determining the bond energy correlators $O^{\gamma}_{ij}=\expval{H^{\gamma}_{ij}}$. The result is plotted in the inset of Fig. \ref{fig3}; bonds with lower energy are denoted by a thick blue line. The lower energy bonds form a chain in NP$_1$. Additionally, a new nematic phase, NP$_3$ emerges for smaller $\alpha$ regime. Unlike NP$_1$, the lower energy bonds form a dimer, implying that the difference between the two NP phases is independent of the ED cluster shape. 

When $\alpha=0$ and $\beta=0$, represented by a red star in Fig. \ref{fig3}, the quantum model belongs to a large class of exactly solvable models\cite{Chulliparambil2020PRB}. At this point, the model can be cast into quadratic form by writing the spin and orbital degrees of freedom in terms of Majorana fermions, $S^{\alpha}_i = - \frac{i}{4} \epsilon^{\alpha \beta \gamma} c_i^\beta c_i^\gamma$ and $T^{\alpha}_i = -\frac{i}{4} \epsilon^{\alpha \beta \gamma} d_i^\beta d_i^\gamma$. Interestingly, ED suggests that the spin liquid phase survives for a large window of $\beta$ along the $\alpha=0$ line denoted by the red line in Fig. \ref{fig3}, since no discontinuities in the ground state energy or its derivatives were observed. This can be contrasted with the classical phase diagram where this line collapses into a single point, suggesting the phase is stabilized by quantum fluctuations. 

When $\alpha=0$ and $\beta \gtrsim 0.19$ indicated by the solid blue line in Fig. \ref{fig3}, a phase transition to another disordered phase is observed. In this phase there is no evidence of long range ordering in both spin {\it{and}} orbital degrees of freedom, unlike the classical model, which orders in the orbital degree of freedom. However, this discrepancy may be due to finite size effects, which requires future studies. 
 
\section{Discussion and Summary } 
As mentioned, when $\alpha = \beta =0$, our Hamiltonian maps to the YL model, but there are two important differences. The Kitaev-type interaction appears on the spin, not on the orbitals, and the orbital interaction is not the Heisenberg type, but is XXZ-like, $T_i^x T_j^x + T_i^z T_j^z - T_i^y T_j^y$. 

In terms of Majorana fermion operators, $S^{\alpha}_i = -\frac{i}{4}\epsilon^{\alpha \beta \gamma} c_i^\beta c_i^\gamma$ and $T^{\alpha}_i = -\frac{i}{4}\epsilon^{\alpha \beta \gamma} d_i^\beta d_i^\gamma$, the model is then written as 
${\cal H} = \frac{1}{8} \sum_{\langle ij \rangle} {\hat u}_{ij} \left( i d_i^x d_j^x + d_i^z d_j^z - d_i^y d_j^y\right)$ where ${\hat u}_{ij} = -i c_i^\gamma c_j^\gamma$.
This implies that it is the orbitals that fractionalize into gapless Majorana fermions in the background of the zero-flux sector composed of the $c$ Majoranas. Note that the $e_g$ orbitals transform like pseudospins, and $T_x$ and $T_z$ carry quadrupolar moments, while $T_y$ carries an octupolar moment;
 $T_x = 1/(2\sqrt{3}) Q_{x^2-y^2}$
$T_y = 1/(3\sqrt{5}) O_{xyz}$
$T_z = 1/(2\sqrt{3}) Q_{3z^2-y^2}$.
A complex fermion operator can then obtained by combining $d_i^z$ and $d_i^x$ Majorana fermions, $f_{i,y} = (d_i^z - i d_i^x)/2$, implying fermionic  octupolar excitations in the ground state.

It is also worth drawing comparisons to models such as the spin-1/2 $K\Gamma$ model in a finite field and the spin-1 bi-linear bi-quadratic Kitaev (BBQ-K) model which exhibit nematic orders bordering the spin liquid region. \cite{Lee2020Natcom,GohlkePRR2020,Rico2023PRB,Mashiko2024PRB} In particular, for the $K\Gamma$ model, as the $\Gamma$ interaction is increased in a finite field along the [111] direction, there is a transition between two nematic phases with the same bond-energy correlators as our $\textrm{NP}_3$ and $\textrm{NP}_1$ phases.  The BBQ-K model also undergoes a transition, at a large enough bi-quadratic interaction, from the Kitaev spin-liquid through a nematic phase before entering a ordered stripy spin phase (see Fig. 3 of \cite{Mashiko2024PRB}). A similar sequence of transitions occurs in our model, as shown in Fig. \ref{fig3}, when $\beta \approx 0.19$.

There are many exciting avenues for future exploration. One possible direction is to identify microscopic mechanisms to tune the system toward the exactly solvable point or to expand its stability in phase space. 
For instance, exchange processes such as cyclic and two-hole exchanges, which are beyond the scope of the current study, will introduce additional symmetry-allowed interactions and may reduce some unwanted interactions. Investigating these interactions and their impacts on the  YL spin-orbital liquid's stability window would be valuable.
In terms of material candidates, $d^9$ ions on a  honeycomb lattice with edge-sharing ligands with strong SOC presents a promising option. $d^7$ (such as Co$^{2+}$) could be another candidate, but due to the strong Hund's coupling,  achieving a low-spin state, one electron in the e$_g$ orbital, while keeping the octahedra crystal field symmetry may be challenging. 
We hope that our theory will inspire further studies on extended models and the search for YL spin-orbital liquid candidates. 

In summary, we present the first microscopic theory that links the extensively studied KK model to the flavoured Kitaev limit, referred to as the YL model. We demonstrate that $d^9$ ions (or $d^7$ ions with sufficiently large cubic crystal field splitting, leading to a low spin state) surrounded by heavy ligands with SOC coupling are positioned near large swaths of nematic phases that engulf an exactly solvable spin liquid point, verified through both classical and quantum simulations. 
Our work serves as a starting point for the search for materials that exhibit this new class of flavored Kitaev physics featuring fermionic octupolar excitations. 

\section{Acknowledgments} 
H.Y.K thanks P. Coleman and A. Tsvelik for introducing the Yao-Lee model and for their stimulating discussions.  
This work is supported by the NSERC Discovery Grant No. 2022-04601. H.Y.K acknowledges support from the Canada Research Chairs Program and 
the Simons Emmy Noether fellowship of the Perimeter Institute,
supported by a grant from the Simons Foundation (1034867, Dittrich).
Computations were performed on the Niagara supercomputer at
the SciNet HPC Consortium. SciNet is funded by: the
Canada Foundation for Innovation under the auspices of
Compute Canada; the Government of Ontario; Ontario
Research Fund - Research Excellence; and the University
of Toronto.

\vspace{2mm}


%

\end{document}


\title{Supplementary Information}
\author{Derek Churchill}
\author{Emily Z. Zhang}
\affiliation{Department of Physics, University of Toronto, Ontario, Canada M5S 1A7}
\author{Hae-Young Kee}
\email[]{hy.kee@utoronto.ca}
\affiliation{Department of Physics, University of Toronto, Ontario, Canada M5S 1A7}
\affiliation{Canadian Institute for Advanced Research, CIFAR Program in Quantum Materials, Toronto, Ontario, Canada, M5G 1M1}
\date{\today}

\newcommand{\nmg}[2]{n_{#1 #2}}
\newcommand{\nmu}[1]{n_{#1 \uparrow}}
\newcommand{\nmd}[1]{n_{#1 \downarrow}}

\newcommand{\cmg}[2]{c_{#1 #2}^{\dag}}
\newcommand{\cmu}[1]{c_{#1 \uparrow}^{\dag}}
\newcommand{\cmd}[1]{c_{#1 \downarrow}^{\dag}}

\newcommand{\amg}[2]{c_{#1 #2}}
\newcommand{\amuu}[1]{c_{#1 \uparrow}}
\newcommand{\amd}[1]{c_{#1 \downarrow}}
\newcommand{\df}{J\sum_{m \neq m'}\cmu{m}\cmd{m}\amd{m'}\amuu{m'}}

\maketitle

\section{I. Derivation of the Effective Model}

We begin this section with a brief review of the local Kanamori-Hubbard Hamiltonian with spin-orbit coupling, which describes the Coulomb interaction for multi-orbital atoms. \cite{georges2013ARP} The interaction is given by
\begin{align} \label{kanamoriHam}
    H_\textrm{int} &= U\sum_{m}\nmu{m}\nmd{m} + U'\sum_{m\neq m'}\nmu{m}\nmd{m'} + \left(U' - J_H\right) \sum_{m < m', \sigma} \nmg{m}{\sigma} \nmg{m'}{\sigma} \\
    \notag & + J_H \sum_{m \neq m'}\cmu{m}\cmd{m'}\amd{m}\amuu{m'} + J_H \sum_{m \neq m'}\cmu{m}\cmd{m}\amd{m'} \amuu{m'} - \lambda {\bf L} \cdot {\bf S} + \epsilon \sum_{m,\sigma}\nmg{m}{\sigma},
\end{align}
where $J_H$ is Hund's coupling, $U$ is the intra-orbital Coulomb interaction, $U'$ is the interorbital Coulomb interaction, $\lambda$ is spin-orbit coupling, and $\epsilon$ is the chemical potential. Below we report the spectrum of this Hamiltonian for the $M$
and $A$ sites restricted to the relevant particle sectors we use in the following subsections.
{\renewcommand{\arraystretch}{1.7}
\setlength{\tabcolsep}{0.1cm}
\begin{center}
    \begin{tabular}{|c c c c c |}
        \hline
        \raggedright M Site  &  &  &   &   \\ [0.5ex] 
        \hline\hline
         & &\underline{2 electron} & & \\  
         degeneracy=6 & & $E_{M,2}=U_M + 2\epsilon_M$   & & \\
         \hline\hline
         & &\underline{1 electron}&& \\  
         degeneracy=4& & $E_{M,3}=\epsilon_M$   & & \\  
         \hline\hline
         & &\underline{0 electron} & & \\
         degeneracy=0& & $E_{M,4}=0$   & & \\
         \hline      
    \end{tabular}
    \captionof{table}{Spectrum of the Eq. \ref{kanamoriHam} Hamiltonian for an M site with $e_g$ orbitals and various fillings. We assume the $e_g$ are well isolated from the $t_{2g}$ orbitals so that spin-orbit coupling is zero in this subspace, and have assumed the Coulomb interaction is much larger than Hund's coupling.}
\end{center}
}

{\renewcommand{\arraystretch}{1.7}
\begin{center}
\setlength{\tabcolsep}{0.1cm}
    \begin{tabular}{|c c c c c |}
        \hline
        \raggedright A site   &  &  &   &   \\ [0.5ex] 
        \hline\hline
         & &\underline{0 hole} & & \\  
         degeneracy=1 & & $E_{A,0}= 15 U_A + 6\epsilon_A$   & & \\
         \hline\hline
         & &\underline{1 hole, $\lambda = 0$}&& \\  
         degeneracy=6& & $E_{A,1}= 10 U_A + 5\epsilon_A$    & & \\  
         & &\underline{1 hole, $\lambda \neq 0$}&& \\  
         degeneracy=2& & $E_{A,1,1}= E_{A,1} + \lambda$   & & \\  
         degeneracy=4& & $E_{A,1,2}= E_{A,1} - \frac{\lambda}{2}$   & & \\
         \hline
    \end{tabular}
    \captionof{table}{Spectrum of the Eq. \ref{kanamoriHam} Hamiltonian for an A site with $p$ orbitals and various fillings. We denote the Coulomb interaction strength on the $A$ site by $U_A$ and spin-orbit coupling by $\lambda$.}
    \end{center}
}

\begin{figure}[h]
    \includegraphics[width=0.35\textwidth]{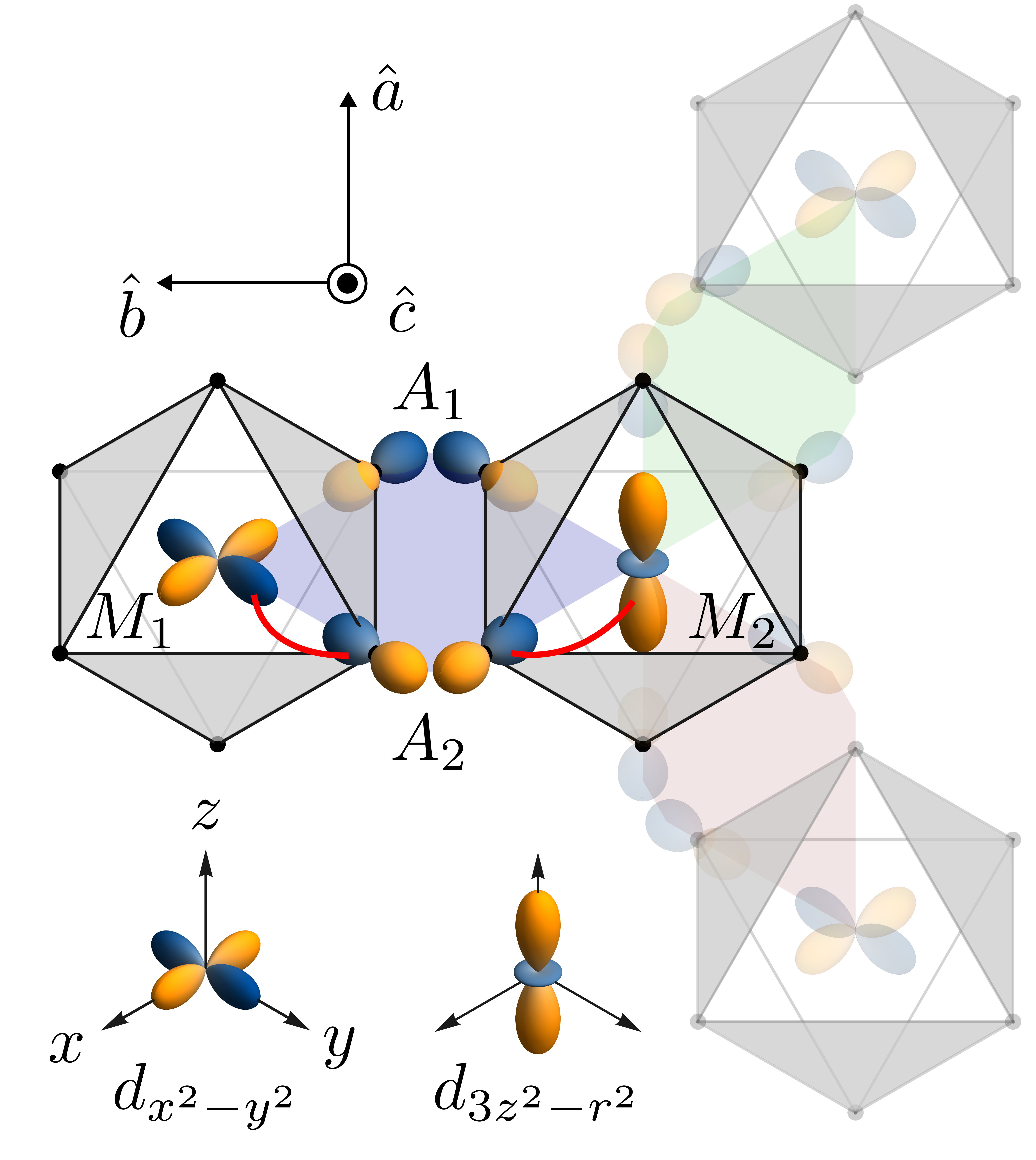}
    \caption{\label{fig1} Schematic of the effective inter-orbital hopping process between $M$ sites via p-orbitals on $A$ sites; the hopping $\propto t_{pd\sigma}$ between $M$ and $A$ sites is represented by red lines. We denote the $x$, $y$ and $z$ bonds by red, green and blue planes which are related by $C_3$ symmetry.}
\end{figure}

We first focus on a $z$-bond, highlighted in Fig. $\ref{fig1}$ and determine an effective hopping between nearest-neighbour sites $M_i$ and $M_j$ by integrating out the intermediate $A$ site p-orbitals. The ground state electron configuration of $A_i$ and $M_i$ ($i=1,2$) is $p^6$ and $e_{g}^1$ respectively. The tight-binding Hamiltonian describing hopping from $A_1$ to $M_1$ is
\begin{align}
    t_{M_1,A_1} = t_{pd\sigma} \sum_{\sigma} \left[-\frac{\sqrt{3}}{2} c^{\dagger}_{M_1, a \sigma} c_{A_1, p_x \sigma} + \frac{1}{2}c^{\dagger}_{M_1, b \sigma} c_{A_1, p_x \sigma} + \textrm{H.c.}\right],
\end{align}
where $t_{pd\sigma}$ is the hopping amplitude.
The Hamiltonians connecting $M_2$-$A_1$, $M_2$-$A_2$ and $M_1$-$A_2$ can be determined by $C_2$ symmetries about the bond center. Assuming that hopping $t_{pd\sigma}$ is much smaller than the charge-transfer gap $\Delta_{pd}\equiv \epsilon_M - \epsilon_A$, the effective tight-binding model from $M_2$ to $M_1$ can be determined at second-order from
\begin{align}
    t^z_{M_1,M_2} = \sum_{(l;j,j_z)} \frac{t_{M_1,A_l} \ket{l;j,j_z}\bra{l;j,j_z} t_{M_2,A_l}}{\Delta_{j}} 
\end{align}
where $(l;j,j_z)$ sums over ligand sites ($l$) and all of its corresponding single hole states ($j$,$j_z$), and $\Delta_{\frac{1}{2}}=\Delta_{pd}+\lambda$ or $\Delta_{\frac{3}{2}}=\Delta_{pd}-\frac{\lambda}{2}$ for the $j=1/2$ and $j=3/2$ states respectively. The resulting effective hopping along the $z$-bond written in second quantization between $M_1$ and $M_2$ sites is 
\begin{align}
    t^z_{M_1,M_2} = i t_{\textrm{eff}} \sum_{\sigma} (-1)^{\sigma} c^{\dagger}_{M_1, a \sigma} c_{M_2, b \sigma} + \textrm{H.c.},
\end{align}
where $t_{\textrm{eff}}= \frac{t_{pd\sigma}^2}{4 \sqrt{3}} \left(\frac{1}{\Delta_{pd}-\frac{\lambda}{2}}-\frac{1}{\Delta_{pd}+\lambda} \right)$, and $\sigma = \pm 1$ for spin up and down respectively. The spin-dependence arises from a combination of the spin-preserving nature of $t_{pd\sigma}$, and the relative signs between anion wavefunction coefficients arising due to time-reversal symmetry.

Using a strong-coupling expansion \cite{Mila2011Spr}, treating $t_{\textrm{eff}}$ as a perturbation to the Coulomb repulsion on the $M_i$ sites, we obtain 
\begin{gather}\label{zbondmodel}
    H_{ij}^z = -J \biggl(S_i \cdot S_j - 2 S_i^z S_j^z - \frac{1}{4}\biggr) \biggl(T^x_i T^x_j +T^z_i T^z_j - T^y_i T^y_j - \frac{1}{4}\biggr),
\end{gather}
for the z bond, where $J \equiv \frac{8 t_{\textrm{eff}}^2}{U_M}$. 

Under $C_3$ rotations about the crystallographic $\hat{c}$ direction, the spin and orbital operators transform as
\begin{gather}
    s^x \rightarrow s^y, \quad s^y \rightarrow s^z, \quad s^z \rightarrow s^x \\
    T^x \rightarrow -\frac{1}{2}T^x - \frac{\sqrt{3}}{2}T^z, \quad T^y \rightarrow T^y, \quad T^z \rightarrow \frac{\sqrt{3}}{2} T^x - \frac{1}{2} T^z
\end{gather}

As a consequence, $T_i^x T_j^x + T_i^z T_j^z$ is invariant under $C_3$ directions about the $\hat{c}$ direction, and the full Hamiltonian is given by
\begin{gather}\label{zbondmodel}
    H_{\textrm{model}} = -J \sum_{\langle i,j \rangle_\gamma} \biggl(S_i \cdot S_j - 2 S_i^\gamma S_j^\gamma - \frac{1}{4}\biggr) \biggl(T^x_i T^x_j +T^z_i T^z_j - T^y_i T^y_j - \frac{1}{4}\biggr).
\end{gather}

\section{II. Numerical Simulation Details}
In this section we describe the details of the Classical Monte Carlo simulations and exact diagonalization in more detail. To encode both degrees of freedom (d.o.f.), we considered a honeycomb bilayer with one lattice representing spin and another representing orbitals. We determined the phase boundaries with both methods by identifying divergent behaviour in $\partial^2 E_{GS}/\partial \alpha^2$ and $\partial^2 E_{GS}/\partial \beta^2$, where $E_{GS}$ is the energy of the groundstate. Phases were  characterized by computing the following structure factors 
\begin{align}
    S(Q)=\frac{1}{N^2}\sum_{ij} \expval{\left(S_i \cdot S_j\right)} e^{-i Q \cdot (r_i-r_j)} \\
    T(Q)=\frac{1}{N^2}\sum_{ij} \expval{\left(T_i \cdot T_j\right)} e^{-i Q \cdot (r_i-r_j)} \\
    ST(Q) = \frac{1}{N^2}\sum_{ij} \expval{\left(S_i \cdot S_j\right)\left(T_i \cdot T_j\right)} e^{-i Q \cdot (r_i-r_j)}.
\end{align}

Classical Monte Carlo simulations were carried out using ClassicalSpinMC.jl (\url{www.github.com/emilyzinnia/ClassicalSpinMC.jl}). We computed the classical phase diagram using a 528 site honeycomb bilayer, generated by a 11x12 triangular lattice with 4 sites per unit cell. Two sites in the unit cell belong to the spin layer, and two belong to the orbital layer, and they are coupled by the quartic interactions between the d.o.f. For each point in the phase diagram, we used an adaptive metropolis simulated annealing algorithm~\cite{Alzate-Cardona2019JPCM} with $10^5$ thermalization steps cooled down to a temperature of $10^{-7}$. During thermalization, $10$ overrelaxation sweeps (also known as microcanonical sweeps)~\cite{Pixley2008PRB} were performed for every Metropolis sweep for faster convergence. The spins were then deterministically updated for $10^6$ sweeps such that they aligned towards the local field~\cite{Janssen2016PRL}. 

\begin{figure}[h]
    \includegraphics[width=0.5\textwidth]{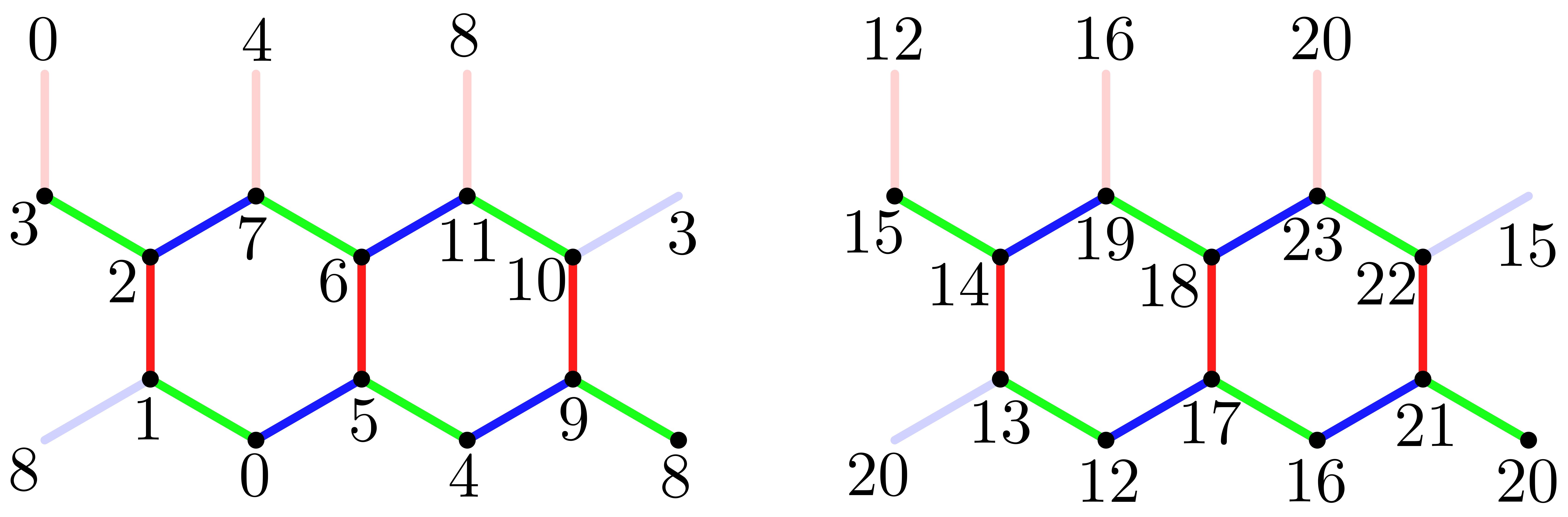}
    \caption{\label{fig7} Exact diagonalization 24-site cluster geometry, with sites labelled 0-23 and x, y, z bonds encoded by red, green, blue lines respectively. Transparent bonds correspond to bonds wrapping around the cluster because of periodic boundary conditions. The left lattice represents the spin degree of freedom and its sites are labelled 0-11. The right lattice represents the orbital degree of freedom with sites $j$ labelled 12-23. Site $j$ on the orbital lattice is identified with site $j$ mod 16 on the spin lattice when considering the quartic interactions in $H_{\textrm{model}}$.}
\end{figure}

Exact diagonalization was performed using the QuSpin python package \cite{Weinberg2019SPP}, with the Lanczos algorithm. We computed the quantum phase diagram using a 24 site honeycomb bilayer (12 sites for each d.o.f.), with periodic boundary conditions as seen in Fig. \ref{fig7}.

\section{III. Comparison of Phases in Classical and Quantum Models}

In this section we present classical spin / orbital configurations and the structure factor for the exactly solvable point, above the exactly solvable point and for the nematic paramagnetic (NP) phases. The intensity in plots for $S(Q)$, $T(Q)$, and $ST(Q)$ are normalized by their maximum values.

\begin{figure}[h]
    \includegraphics[width=0.6\textwidth]{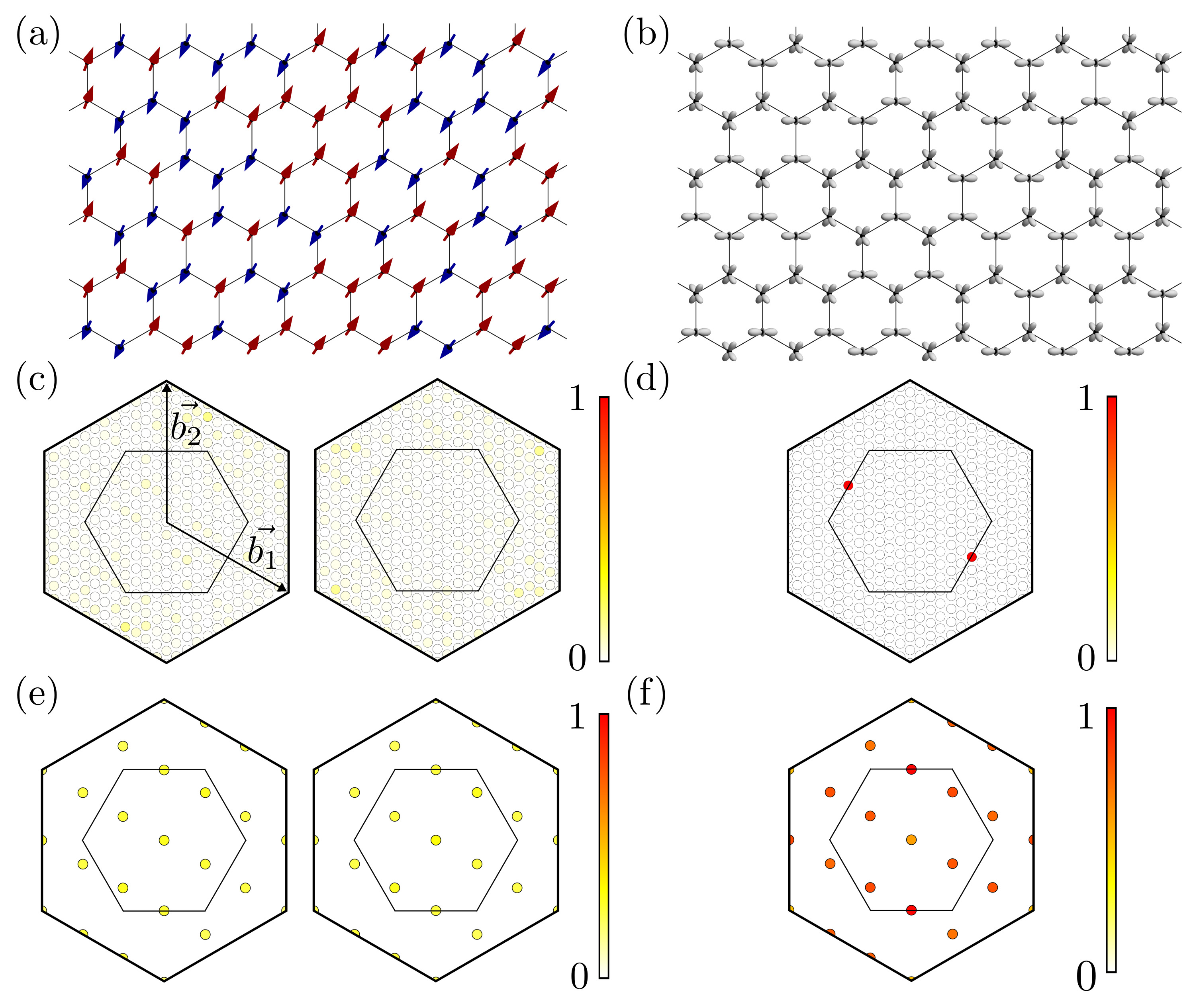}
    \caption{\label{fig2} Classical (a) spin and (b) orbital configuration for the neumatic NP$_1$ phase, $(\alpha,\beta)=(0.5,0.004)$. Classical $S(Q)$ (left) and $T(Q)$ (right) are displayed in (c) and indicate no long range ordering; however there are long-range correlations between d.o.f as shown in the plot of $ST(Q)$ in panel (d). We plot the quantum analogs obtained from exact diagonalization in (e) and (f), respectively. We note that the peaks are much more broad due to finite size effects, but still appear at the $M$ point.}
\end{figure}

\begin{figure}[h]
    \includegraphics[width=0.6\textwidth]{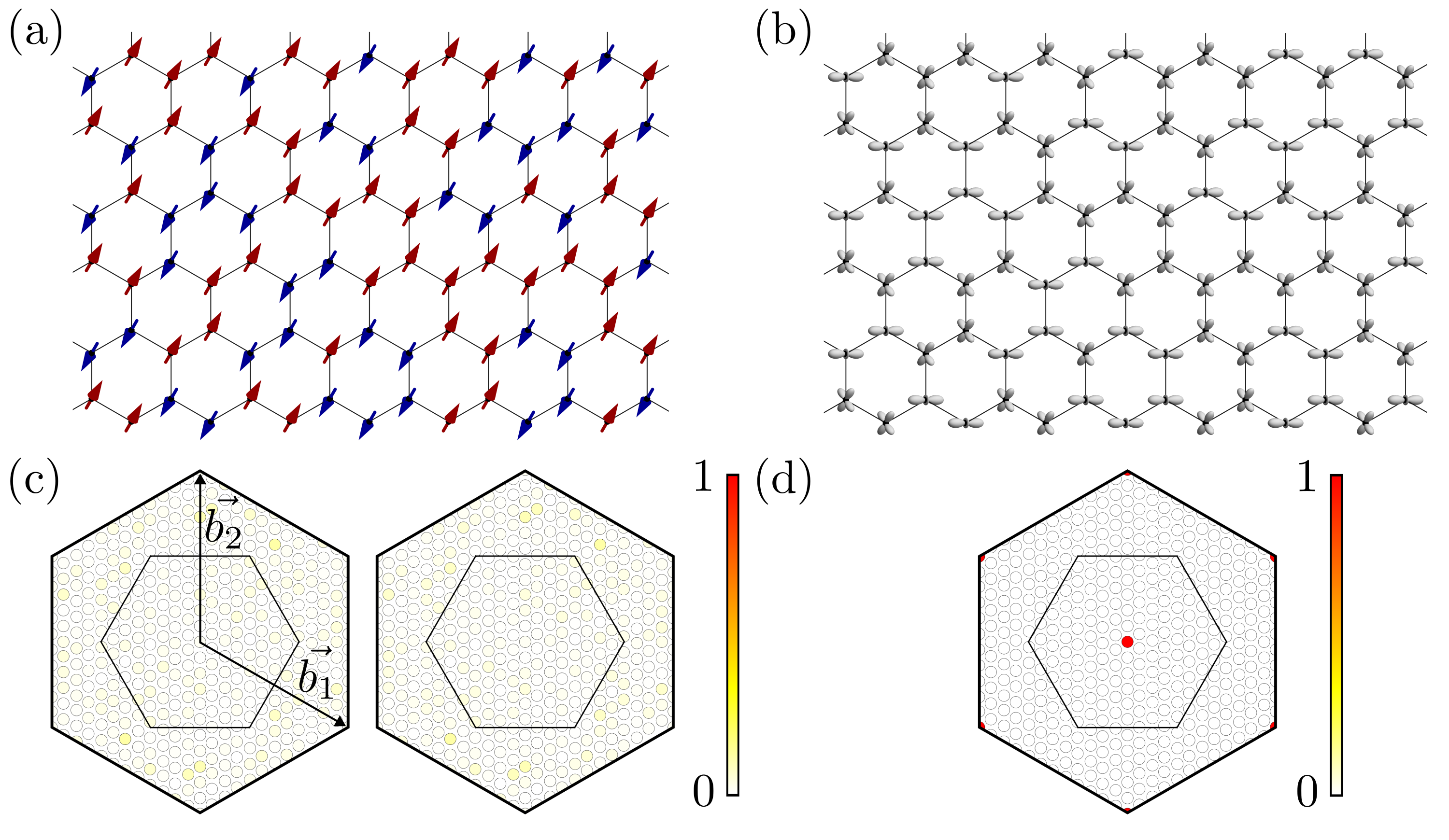}
    \caption{\label{fig3} Classical (a) spin and (b) orbital configuration for the neumatic NP$_2$ phase, $(\alpha,\beta)=(2.5,0.004)$. The classical $S(Q)$ (left) and $T(Q)$ (right) are displayed in (c). The $ST(Q)$ plotted in panel (d) has a peak at the $\Gamma$ point, indicating FM correlations between the d.o.f.. This phase does not appear in the quantum phase diagram in the parameter regime we checked.}
\end{figure}

\begin{figure}[h]
    \includegraphics[width=0.53\textwidth]{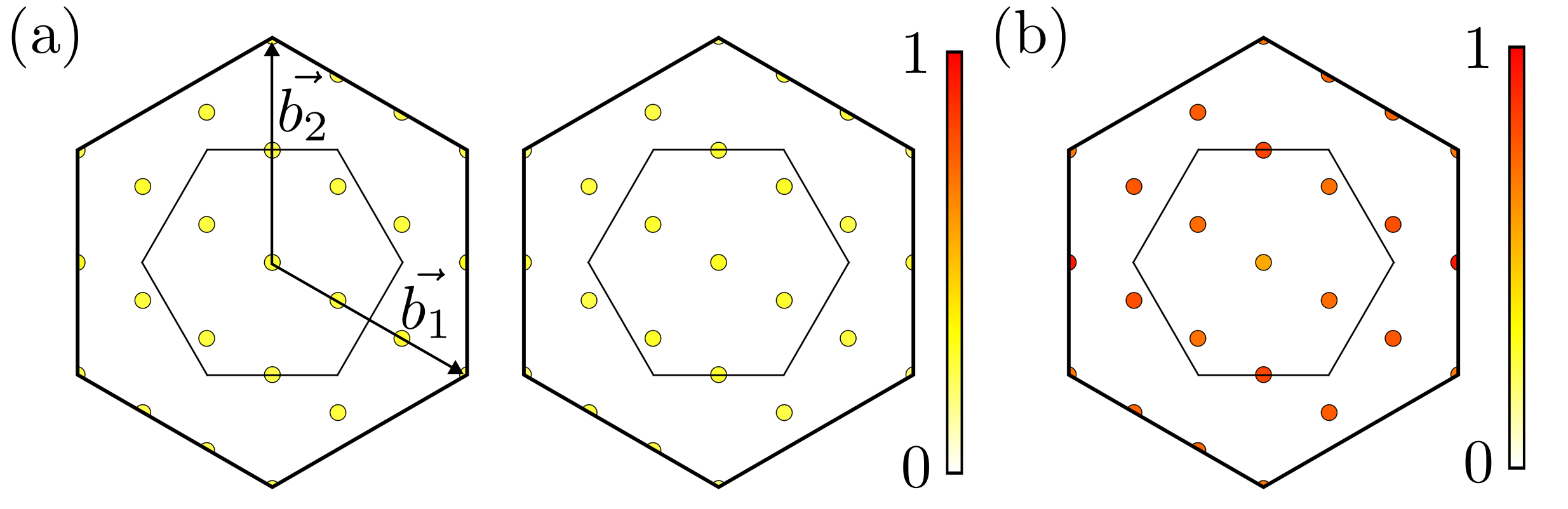}
    \caption{\label{fig4} (a) The $S(Q)$ (left) and $T(Q)$ (right) structure factor and (b) $ST(Q)$ for the NP$_3$ phase in the quantum phase diagram. This phase does not appear as the ground state in the classical phase diagram.}
\end{figure}

\begin{figure}[h]
    \includegraphics[width=0.6\textwidth]{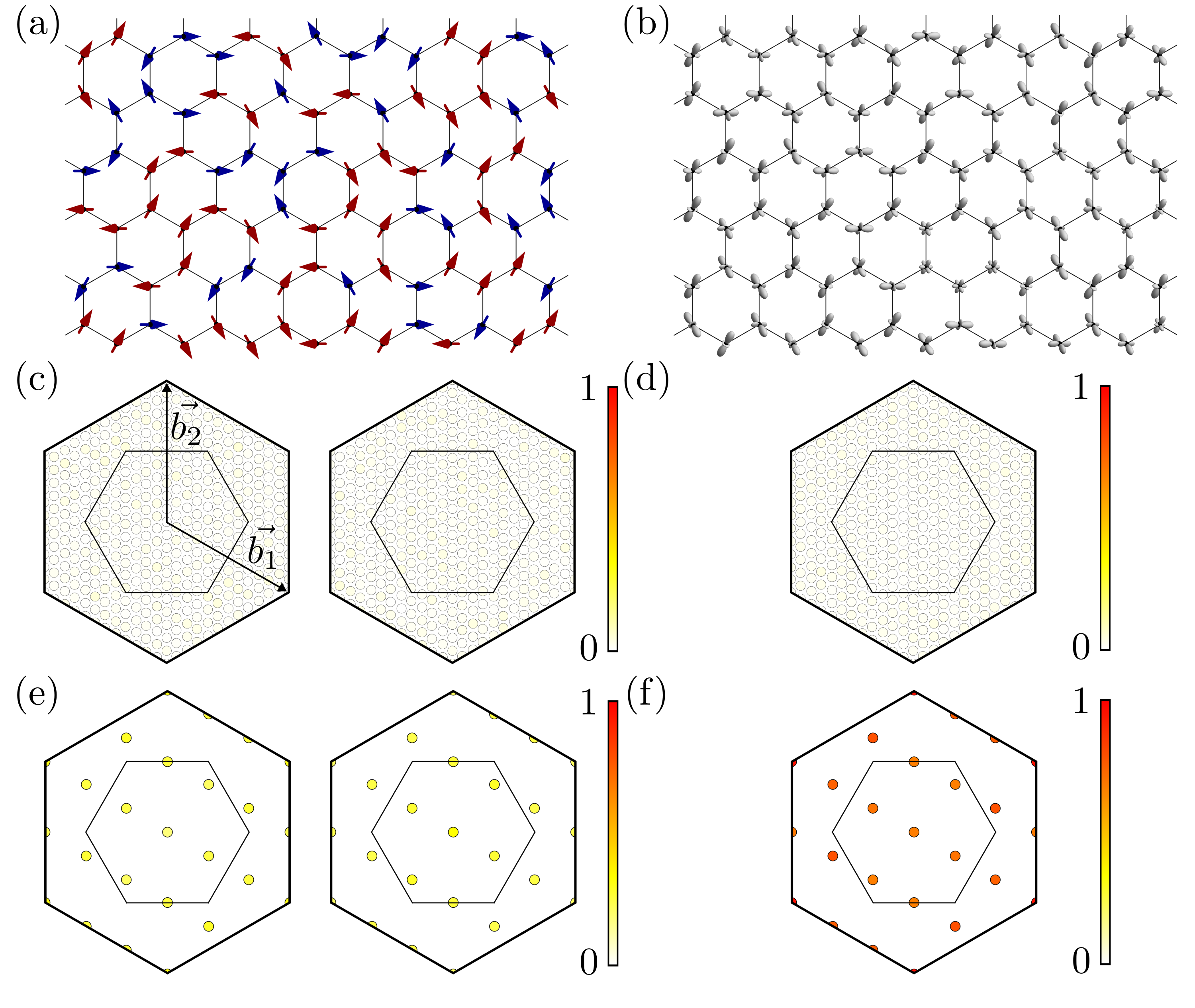}
    \caption{\label{fig5} Classical (a) spin and (b) orbital configuration at the exactly solvable point, $(\alpha,\beta)=(0.0,0.0)$. Classical $S(Q)$ (left) and $T(Q)$ (right) are displayed in (c) and indicate no long range ordering. (d) Classical $ST(Q)$ shows no long-range correlations. Quantum $S(Q)$ (left) and $T(Q)$ (right) in panel (e) confirm the classical results. The spin-orbital correlator shows very broad peaks in contrast with the classical case, but is most likely due to finite size effects.}
\end{figure}

\begin{figure}[H]
    \centering{\includegraphics[width=0.6\textwidth]{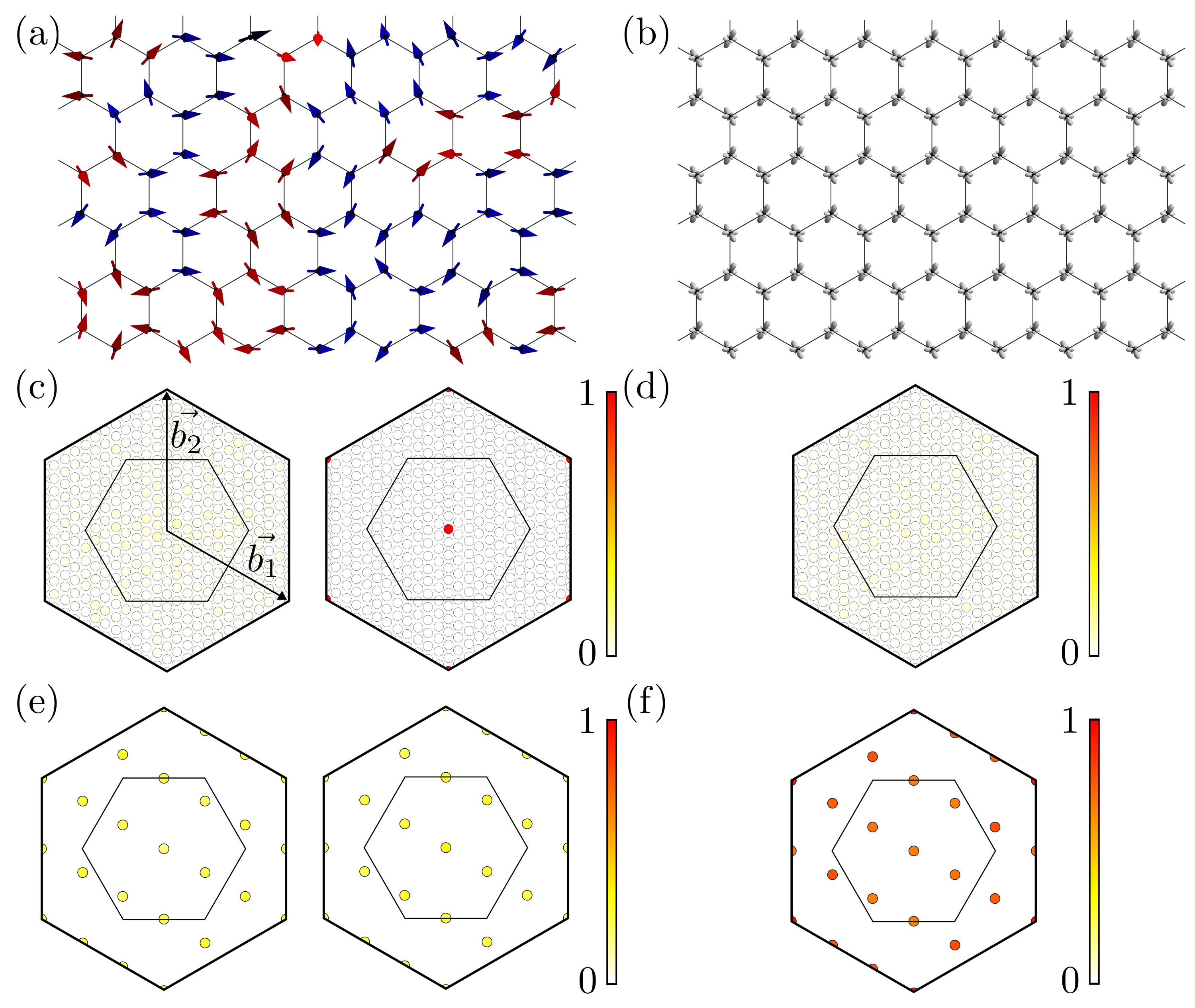}}
    \caption{\label{fig6} Classical (a) spin and (b) orbital configuration above the exactly solvable point,  $(\alpha,\beta)=(0.0,0.05)$. The $S(Q)$ (left) and $T(Q)$ (right) structure factors are presented in (c) and $ST(Q)$ is plotted in (d). In contrast to the exactly solvable point, this phase has long-range orbital ordering due to the small orbital bilinear term which quickly destabilizes the classical spin-orbital phase. The quantum analogs are shown in panels (e) and (f), respectively. The broad peaks which appear in $ST(Q)$ in the quantum model are likely due to finite size limitations of exact diagonalization.}
\end{figure}